\documentclass{aa}  

\usepackage{color}
\usepackage{graphicx}
\usepackage{txfonts}
\usepackage{hyperref}
\newcommand{\de}{\text{d}}
\newcommand{\Msun}{\text{M}_{\odot}}
\newcommand{\kmsec}{\text{km}\,\text{s}^{-1}}

\newcommand{\kpc}{\text{kpc}}
\newcommand{\pc}{\text{pc}}

\newcommand{\Msunppcsquare}{\Msun\pc^{-2}}

\newcommand{\update}[1]{#1}

\begin{document}

   \title{First spiral arm detection using dynamical mass measurements of the Milky Way disk}


   \author{Axel Widmark
          \inst{1}
          \and
          Aneesh P. Naik
          \inst{2}
          }

   \institute{Oskar Klein Centre for Cosmoparticle Physics, Stockholm University, Alba Nova, Stockholm, SE-106 91, Sweden\\
              \email{axel.widmark@fysik.su.se}
         \and
             Institute for Astronomy, University of Edinburgh, Royal Observatory, Blackford Hill, Edinburgh, EH9 3HJ, UK\\
             \email{aneesh.naik@roe.ac.uk}
             }

   \date{Received XX XX, XXXX; accepted XX XX, XXXX}

 
  \abstract{
  We apply the vertical Jeans equation to the Milky Way disk, in order to study non-axisymmetric variations in the thin disk surface density. We divide the disk plane into area cells with a 100~pc grid spacing, and use four separate subsets of the \emph{Gaia} DR3 stars, defined by cuts in absolute magnitude, reaching distances up to 3~kpc. The vertical Jeans equation is informed by the stellar number density field and the vertical velocity field; for the former, we use maps produced via Gaussian Process regression; for the latter, we use Bayesian Neural Network radial velocity predictions, allowing us to utilize the full power of the \emph{Gaia} DR3 proper motion sample. For the first time, we find evidence of a spiral arm in the form of an over-density in the dynamically measured disk surface density, detected in all four data samples, which also agrees very well with the spiral arm as traced by stellar age and chemistry. We fit a simple spiral arm model to this feature, and infer a relative over-density of roughly 20~\% and a width of roughly 400~pc. We also infer a thin disk surface density scale length of 3.3--4.2~kpc, when restricting the analysis to stars within a distance of 2~kpc.
  }


   \maketitle
%

\section{Introduction}\label{sec:intro}

In recent years, observations from the \textit{Gaia} satellite \citep{Gaia2016mission} have revolutionised our understanding of the structure, history and composition of our Galaxy. \textit{Gaia} has measured the positions and motions of more than a billion stars, enabling unprecedented analyses of Milky Way stellar dynamics and stellar population properties.

As an example of this progress, many recent studies have sought a better understanding of the spiral structure of the Milky Way \citep{Xu2018,Shen2020}, which can yield insights into the Galaxy's formation history. The spiral arms have been probed using various tracers, such as hydrogen gas \citep{Levine2006}, masers \citep{Xu2013,Reid2014,Reid2019}, and young stars \citep{Poggio2021,Castro-Ginard2021,Lin2022,GaiaDrimmel2023,Gaia2023RecioBlanco}. Different studies report slightly different arm morphologies, in terms of pitch angle and spatial extent (e.g., comparing \citealt{Reid2019} and \citealt{Poggio2021}; sometimes such differences can, at least in part, be explained by intrinsic differences between tracers, as discussed by \citealt{Miyachi2019}). The spiral arms have also been associated with various kinematic signatures \citep{Eilers2020,WidmarkGP,Khoperskov2022,Martinez-Medina2022,Palicio2023}.

Despite the aforementioned advances in data precision and depth, the Milky Way's spiral structure has not previously been detected in a dynamical mass measurement.
\footnote{\update{\cite{McGaugh2019} uses the Milky Way rotation curve as observed in gas (21 cm and CO) to infer the disk surface density as a function of Galactocentric radius, assuming the radial acceleration relation \citep[RAR;][]{McGaugh2016, Lelli2017}. While they do see variations at the locations of the spiral arms, these variations are rather large. For example, their solar radius value is under-dense with a total disk surface density below $40~\Msunppcsquare$; other studies report roughly $55~\Msunppcsquare$ for the baryonic disk components of the solar neighbourhood, both in dynamical mass measurements as well as direct observation of stars and gas \citep[e.g.,][]{Read2014,McKee2015,Schutz2018,Widmark-sub-regions-DR2,Widmark-spiral-II,Nitschai2021}. It seems likely that the density variations inferred by \cite{McGaugh2019} are biased by uncontrolled systematics, probably in the translation from gas terminal velocity curve to rotation curve. Hence, we characterise their observed correlation with the spiral arms as a kinematic signal, rather than a robust and accurate dynamical mass measurement.}}
To detect such variations in the disk surface density requires precise measurements in distant disk regions. This is challenging, mainly because of stellar crowding and dust extinction that cause strong selection effects and other data biases that are difficult to model. Data incompleteness is especially severe for the radial (i.e., line-of-sight) velocity measurements, which are only available for a bright sub-sample of stars.

In the present work, we use state-of-the-art data processing techniques to overcome these obstacles, and perform vertical Jeans analysis of the thin disk out to a distance of 3~kpc from the Sun. Vertical Jeans analysis was formulated a century ago by \cite{Jeans1922}, and remains a widely used probe of the dynamical properties of the Solar neighbourhood, such as the local density of dark matter \citep{Sivertsson2018, Salomon2020, Guo2020, Salas2021, Widmark-sub-regions-DR2}. Under this technique, one assumes dynamical equilibrium and that stellar motions in the directions parallel and perpendicular to the disk plane can be approximately decoupled. Then, a stellar tracer population's number density distribution and vertical velocity distribution are interrelated via the vertical gravitational potential.

Using the \emph{Gaia} DR3 catalogue, we construct four separate data samples of bright stars, defined by cuts in absolute magnitude. We divide the disk plane into a square grid with a 100~pc spacing, and apply the vertical Jeans analysis to each such area cell and data sample separately. For the two key ingredients of vertical Jeans analysis, we do the following. For the stellar number density field, we use the 3d maps produced by \cite{WidmarkGP} using Gaussian Processes, based on \emph{Gaia} and \texttt{StarHorse} spectro-astrometric distances \citep{starhorse}. For the vertical velocity fields, we use \emph{Gaia} measurements supplemented by Bayesian Neural Network predictions of radial velocity (i.e., line-of-sight velocity) from \cite{Naik_RVs_DR3}.\footnote{The \cite{Naik_RVs_DR3} catalogue of radial velocity predictions is available via the \emph{Gaia} mirror archive, hosted by the Leibniz-Institut f\"ur Astrophysik Potsdam (AIP); DOI: \href{https://doi.org/10.17876/gaia/dr.3/110}{\texttt{10.17876/gaia/dr.3/110}}.} These radial velocity predictions are especially useful for our application, because the radial velocity is a sub-dominant component of the vertical velocity in distant regions of the disk, allowing us to make full use of the highly informative \emph{Gaia} proper motion sample. These data products, in combination, allow us to make a detailed map of how the surface density varies in the directions parallel to the Galactic plane. 

For the first time, we find evidence for the imprint of spiral structure in the dynamically measured vertical gravitational potential of the Milky Way disk. In all of our four data samples, we see the same non-axisymmetric over-density, which also agrees very well with the Local Arm as traced by stellar age and chemistry (e.g.~\citealt{Poggio2021}).

\section{Method}\label{sec:method}

\subsection{Coordinate system and data samples}\label{sec:data_samples}

We use Solar rest-frame Cartesian coordinates $\boldsymbol{X} = (X,Y,Z)$ pointing in the directions of the Galactic centre, Galactic rotation, and Galactic north. The vertical coordinate in the disk frame is equal to
\begin{equation}
    z = Z + Z_\odot,
\end{equation}
where $Z_\odot$ is the height of the Sun with respect to the Galactic disk mid-plane. The vertical velocity in the Solar rest-frame is $W \equiv \de Z / \de t$. The Galactocentric radius is given by
\begin{equation}
    R = \sqrt{(R_\odot - X)^2 + Y^2},
\end{equation}
where we take $R_\odot = 8.2~\kpc$ \citep{mcmillan2016mass}. The Galactocentric angle $\phi$ fulfils
\begin{equation}
    \sin \phi = \frac{Y}{R}, \quad
    \cos \phi = \frac{R_\odot-X}{R}.
\end{equation}

We construct four separate stellar samples by making cuts in \emph{Gaia} $G$-band absolute magnitude, according to $M_G \in (0, 1],\, (1, 2],\, (2, 3]$, and $(3, 4]$ (the same as in \citealt{WidmarkGP}). We divide the $(X,Y)$-plane into a square grid with a cell spacing of 100~pc, for the spatial volume within $\sqrt{X^2+Y^2}<3~\kpc$.

\subsection{Vertical Jeans analysis}\label{sec:jeans}

The vertical Jeans equation gives
\begin{equation}\label{eq:vertical_jeans}
    \frac{1}{n}\frac{\partial (n \sigma^2_W)}{\partial Z} + \frac{\partial \Phi}{\partial Z} = 0,
\end{equation}
where $\Phi$ is the gravitational potential, $n$ and $\sigma_W^2$ are the number density and vertical velocity variance of a stellar tracer population, and $Z$ is the direction perpendicular to the Galactic plane.

In the equation above, we have made a few simplifying assumptions. In particular, we have assumed dynamical equilibrium and dropped a term representing a coupling between radial and vertical motions (the `tilt' term), which gives a sizable contribution only at heights closer to 1~kpc \citep{Read2014}.

\subsection{Stellar number density field}
\label{sec:nu_field}

For the stellar number count density, we use the results from \cite{WidmarkGP}, where the stellar number density fields of the four data samples were modelled with Gaussian Process (GP) regression. This was based on \texttt{StarHorse} spectro-astrometric distances \citep{starhorse}, using data from \emph{Gaia} DR3 \citep{GaiaDR3}, Pan-STARRS1 \citep{Scolnic2015}, SkyMapper \citep{Casagrande2019}, 2MASS \citep{Skrutskie2006}, and AllWISE \citep{Marrese2019}.

\update{As discussed in Sect.~\ref{sec:intro}, selection effects are a significant obstacle to dynamical mass measurements in distant regions of the disk. The \emph{Gaia} selection function is highly complex, and depends on brightness limits, stellar crowding and the \emph{Gaia} scanning law \citep{Boubert-II,everall2022,rybizki2021,CantatGaudin2022}.
Furthermore, the stellar number density field can be biased by open clusters and dust extinction. All these issues were carefully considered and mitigated in \cite{WidmarkGP}.
To avoid issues with the bright and dim end,
they limited the $G$-band apparent magnitudes to the range of roughly 6--17~mag, where the \emph{Gaia} catalogue is very close to complete, with deviations of a few percent at most \citep{everall2022,CantatGaudin2022}.
Furthermore,} spatial volumes were carefully masked to avoid open clusters and incompleteness issues associated with dust extinction. By using GPs, the stellar number density field was modelled as a smooth and differentiable function. This approach was also fairly model-independent, and did not impose constraints such as Galactic axisymmetry.

The GP regression used correlation lengths of $(l_X,l_Y,l_Z)=(300,300,100)~\pc$, which imposes a degree of smoothness in all three spatial directions. In this manner, even though the area cells have a 100~pc spacing in $(X,Y)$-plane, the number density field is informed by a larger area. This is especially useful when a larger fraction of an area cell is masked (e.g., due to an open cluster); in that case, its number density field is still informed by the data in its surrounding spatial volume. However, in order to avoid area cells that are very poorly sampled, we mask area cells that have an average effective volume fraction smaller than 0.5 for $|Z|<250~\pc$.

We want to retain the useful properties of GP regression discussed above. However, there are time-varying perturbations to the stellar number population (which the GP regression encapsulates); such perturbations can create a biases in dynamical mass measurements, since they are not in a steady state \update{(see Appendix~\ref{app:biases} and Sect.~\ref{sec:discussion} for further details)}. Therefore, we fit an analytic function to the stellar number density that was obtained via GP regression, separately for each data sample and area cell, assuming a functional form which is mirror symmetry and monotonically decreasing with height. This analytic function is a mixture model of three disk components, according to
\begin{equation}\label{eq:nu_of_z}
    \hat{n}(Z \, | \, a_i, h_i, Z_\odot) 
    = \sum_{i=1}^3 a_i \times \text{cosh}^{-2}\left( \frac{Z+Z_\odot}{h_i} \right),
\end{equation}
where $a_i$ are the disk component normalisations, $h_i$ are the scale heights, and $Z_\odot$ is the height of the Sun with respect to the disk mid-plane. This functional form is flexible enough to faithfully model the vertical number density profile of our stellar tracer populations. We use wide flat box priors and the GP regression uncertainties in our fit.

\begin{figure*}
    \centering
    \includegraphics[width=0.8\textwidth]{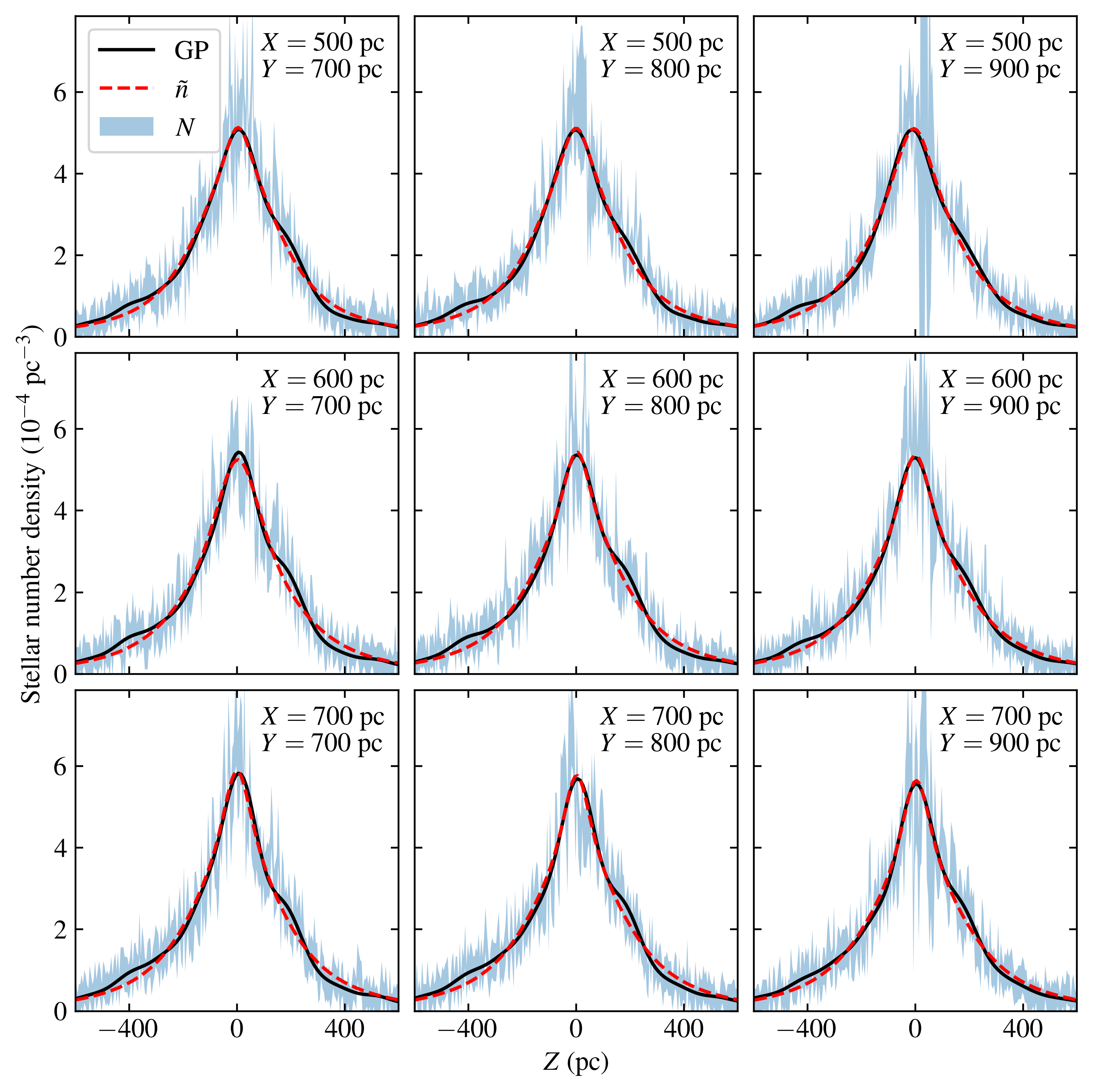}
    \caption{Stellar number density distribution for a few area cells of the data sample with $2 < M_G \leq 3$. Each panel is labelled by the area cell's mid-point $(X,Y)$-coordinates. The function obtained by GP regression is shown in solid black; the fitted symmetric function ($\hat{n}$) of Eq.~\ref{eq:nu_of_z} is shown in dashed red; the raw number count ($N$) with its associated Poisson noise uncertainty is shown as a 1$\sigma$ band.}
    \label{fig:nu_example}
\end{figure*}

In Fig.~\ref{fig:nu_example}, we show an example of the stellar number density distribution, in terms of the underlying data counts, GP regression, and symmetric analytic function. We do so for a small area of nine neighbouring area cells of the $2 < M_G \leq 3$ data sample. We chose this particular spatial volume in order to illustrate a few important features. In the top right panel, the raw data number count has a very large uncertainty, due to open cluster masking. The GP curve is still inferred there, informed by its surrounding spatial volume. In all panels, the stellar over-densities at roughly $Z=200~\pc$ and $Z=-400~\pc$ (and their mirrored under-densities) are projections of the phase space spiral (see figure 7 in \citealt{WidmarkGP}). Because the phase-space spiral is a time-varying structure, it constitutes a bias to Jeans analysis, which we average away by fitting the symmetric and analytic function $\hat{n}$.

\subsection{Vertical velocity field}
\label{sec:w_field}

\begin{figure*}
    \centering
    \includegraphics[width=0.8\textwidth]{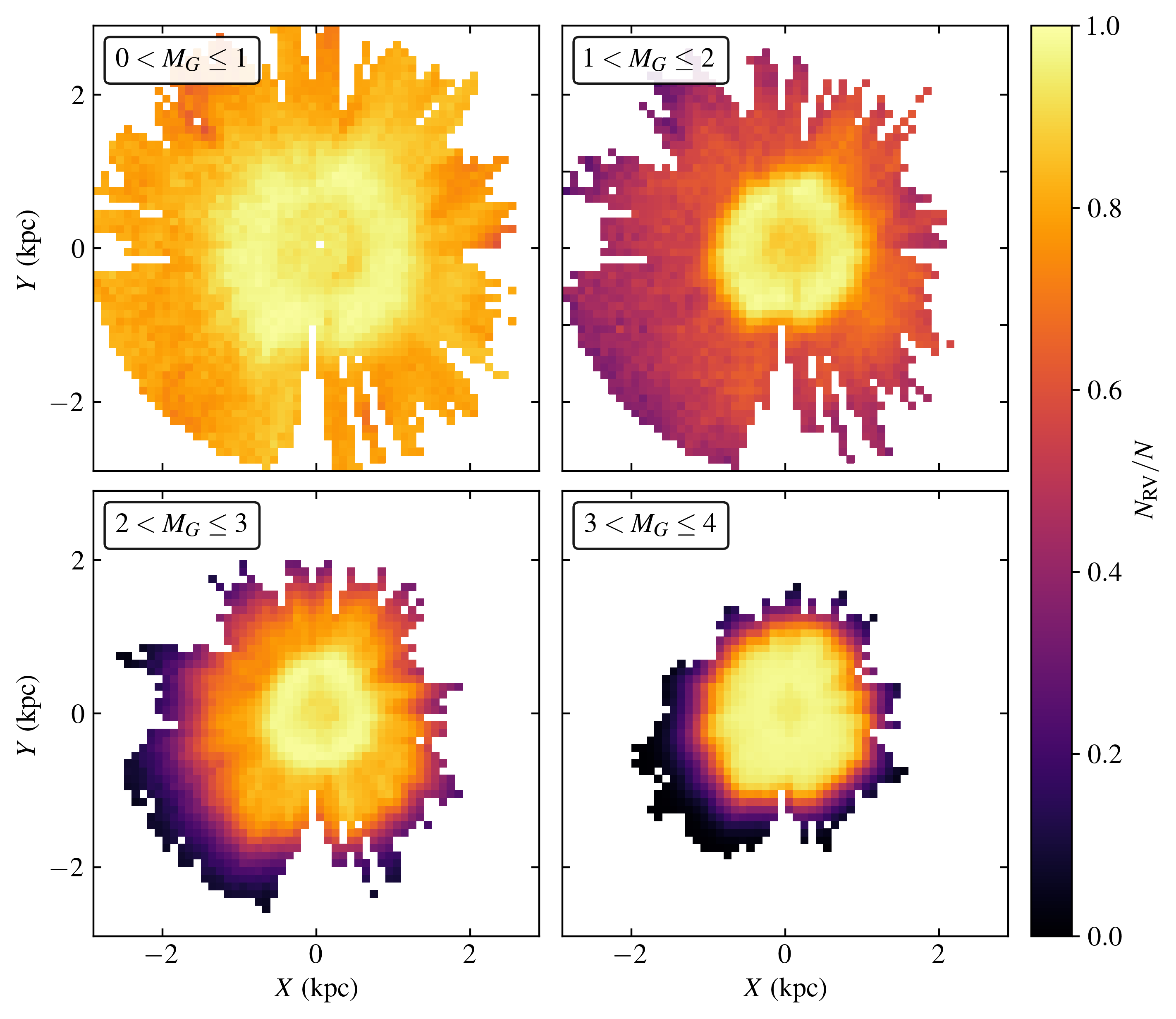}
    \caption{\update{Fraction of stars with radial velocity measurements across the $100\times100$~pc area cells used in our analysis. The four panels represent the four absolute magnitude samples, as labelled.}}
    \label{fig:RV_counts}
\end{figure*}

For the vertical velocity field, we use all \textit{Gaia} DR3 stars in the unmasked $(X,Y)$ area cells, subdivided into the four data samples. We remove possible (probability $>0.1$) members of the open clusters tabulated by \cite{Hunt2023} and stars with \texttt{StarHorse} distance precisions worse than 10\%. This leaves 23\,551\,383 stars overall: 3\,262\,301, 2\,736\,077, 5\,710\,148, 11\,842\,857 in each of the four magnitude bins, from brightest to faintest.

Of these stars, around 60\% have radial velocity measurements in \textit{Gaia} DR3 (ranging from 85\% in the brightest sample to 50\% in the faintest). \update{The spatial distribution of these measurements is shown in Fig.~\ref{fig:RV_counts}, showing the fraction of stars with radial velocity measurements in each of our area cells and for each magnitude sample. At all magnitudes, the stars within approximately 1~kpc are very well covered by radial velocity measurements. Beyond this distance, the radial velocity coverage decreases appreciably with brightness.} For the remainder of the stars \update{without radial velocity measurements}, we use the Bayesian radial velocity predictions of \cite{Naik_RVs_DR3}. These prediction posterior distributions are generated using Bayesian neural networks (BNNs), trained on the subset of DR3 stars with radial velocities \citep{Naik_RVs_preDR3, Naik_RVs_DR3}. In effect, these BNNs learn a latent representation of the stellar phase space distribution, convolved with any uncertainty due to a lack of training data. As a result, these prediction distributions can be rather wide, with typical widths in the range 20-30~$\kmsec$. 

\update{These velocity predictions have been validated at several stages. First, \citet{Naik_RVs_preDR3} used a BNN trained on \textit{Gaia} DR2 to produce a set of blind predictions for radial velocity measurements that would subsequently be released in \emph{Gaia} DR3. The predictions were then compared with the measurements in \citet{Naik_RVs_DR3}, and found to have been very successful: the prediction distributions were not only consistent with the measurements, but they also exhibited appropriate levels of confidence, with inflated prediction uncertainties for stars outside the footprint of the training data. \citet{Naik_RVs_DR3} then repeated the exercise and trained a new model on the DR3 data. This newer model was shown to be highly successful in validation tests with DR3 radial velocities excluded from the training data, and also with stars from other spectroscopic surveys. The model was then used to generate a publicly available catalogue of radial velocity predictions for 185 million stars still lacking radial velocity measurements in \textit{Gaia} DR3. It is these tabulated predictions we use in this work.}

\update{\citet{Naik_RVs_DR3} suggest a natural approach for incorporating these predictive posterior distributions into a wider analysis: \textit{multiple imputation} \citep{Little2014, Gelman2004}. Here, one draws $N$ random realisations of the predictive posterior for each 5D star, to construct $N$ `complete' datasets (or `imputations'), then conducts the remainder of the analysis with each imputation separately to infer the parameters of interest. The variation in the parameters of interest across the imputations gives the uncertainty due to the missing data. This is a well-motivated technique for dealing with missing data, as it is formally equivalent to performing a single joint inference over both the missing data and the parameters of interest. In other words, `Bayesian shrinkage' occurs only once for a given set of observations. In the present work, we construct 20 such imputations (i.e, we generate 20 realisations from the prediction posterior distribution of each star), then conduct the remainder of our analysis separately for each imputation.}


For each imputation, data sample, and area cell, we convert the measured phase space coordinates (sky positions, \texttt{StarHorse} distances, proper motions, and radial velocities) to the heliocentric Cartesian coordinate system described above. We then fit a parametric model describing $\sigma^2_W$ as a function of $Z$, using the data of the area cell as well as all adjacent unmasked cells (effectively smoothing our results over a 300-by-300 pc area):
\begin{equation}\label{eq:sigma_of_z}
    \hat{\sigma}^2_W(Z) = 
    \sigma^2_0 + \sigma^2_1 \ln\left[1 + \left(\frac{Z+Z_\odot}{H_1}\right)^2 \right] + 
    \sigma^2_2 \ln\left[1 + \left(\frac{Z+Z_\odot}{H_2}\right)^2 \right],
\end{equation}
where $\sigma^2_{\{0,1,2\}}$ and $H_{\{1, 2\}}$ are dimensionful free parameters in the fit. The quantity $Z_\odot$ represents the height of the Sun with respect to the disk mid-plane; this is not taken as a free parameter but as a constant, given by the fit of the stellar number density model in that area cell. We fit the velocity variance model (Eq.~\ref{eq:sigma_of_z}) to the data $\mathcal{D}$ (comprising stellar heights $Z$, vertical velocities $W$, and observational vertical velocity uncertainties $\sigma_\mathrm{obs}$ propagated from proper motion and distance measurements) by maximising a Gaussian log-likelihood,
\begin{equation}
    \ln \mathcal{L}(\mathcal{D}\, | \, \overline{W}, \sigma^2_{\{0,1,2\}}, H_{\{1, 2\}}) =
    -\frac{1}{2}
    \sum_i \left[
    \frac{(W_i - \overline{W})^2}{\sigma_i^2}
    + \ln \left(2\pi\sigma_i^2\right)
    \right],
\end{equation}
where the index $i$ runs over individual stars, and
\begin{equation}
    \sigma_i^2 \equiv \sigma_{\mathrm{obs},i}^2+\hat{\sigma}^2_W (Z_i | \sigma^2_{\{0,1,2\}}, H_{\{1, 2\}}).
\end{equation}
The mean vertical velocity $\overline{W}$ is a sixth free parameter, in addition to the five free parameters $\sigma^2_{\{0,1,2\}}, H_{\{1, 2\}}$ entering the dispersion model.

\begin{figure*}
    \centering
    \includegraphics[width=0.8\textwidth]{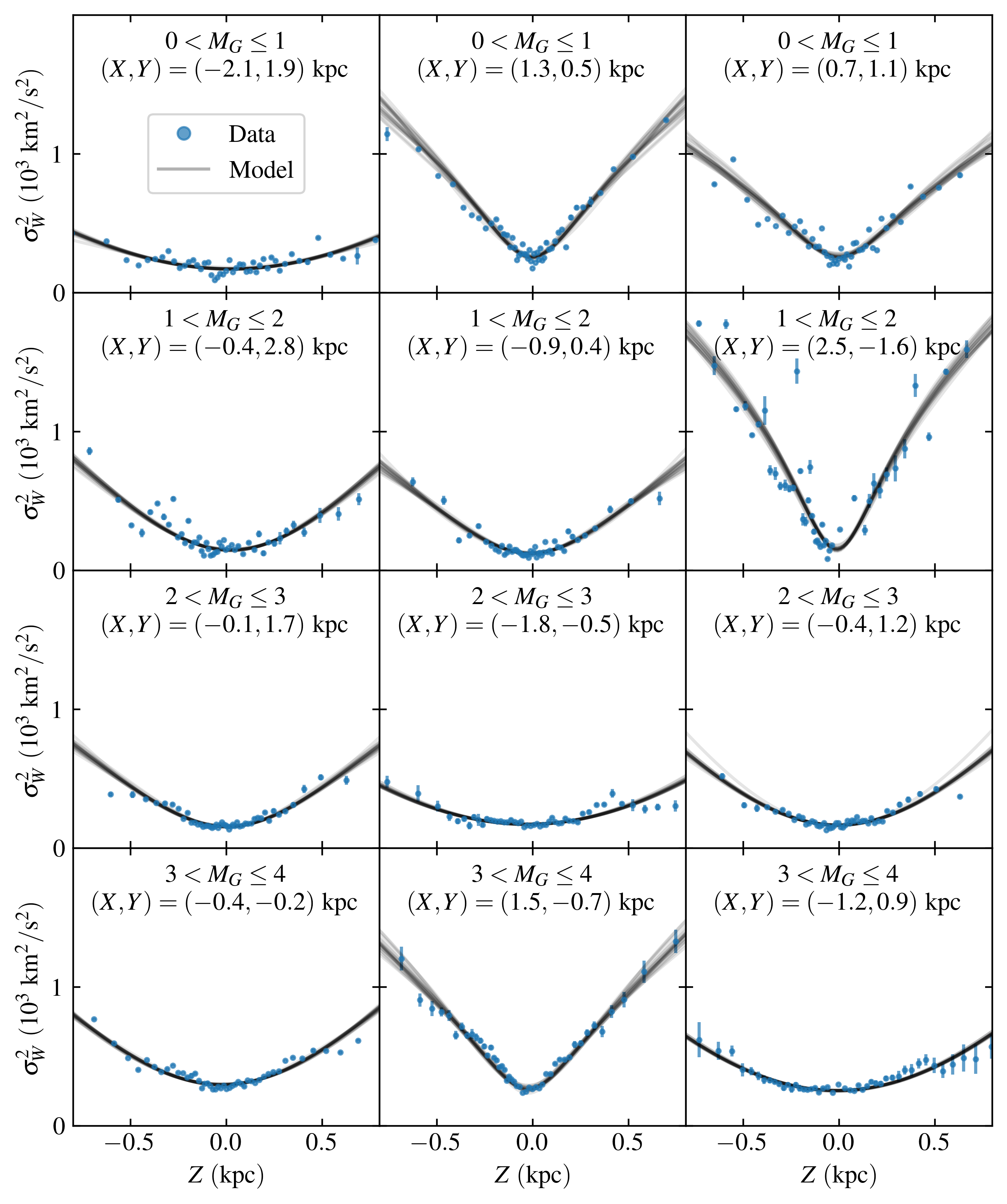}
    \caption{Examples of vertical velocity dispersions and the best-fitting models (Eq.~\ref{eq:sigma_of_z}). The four rows correspond to the four magnitude samples, while the three panels within each row are randomly chosen cells in $(X,Y)$-plane. The yellow points represent the measured velocity dispersions in adaptively spaced $Z$ bins, while the solid lines show the best-fitting models. Note that the model is not fit to these binned dispersions, but to the individual stellar velocities. In each case, 20 sets of data and 20 models are shown, corresponding to the 20 random imputations of the dataset as described in the main text.}
    \label{fig:disp_fit}
\end{figure*}

Figure~\ref{fig:disp_fit} shows a few examples of fitted velocity dispersion profiles. For each of the four data samples, we choose three random area cells and plot the best-fit model according to Eq.~\ref{eq:sigma_of_z}, as well as the measured velocity dispersions in adaptively spaced vertical bins. In each case, we individually plot the data and model for each of the 20 imputations. It is clear from this figure that our model (Eq.~\ref{eq:sigma_of_z}) effectively captures the overall shape of the dispersion profiles despite the observed variability in these shapes, while at the same time foregoing smaller-scale features arising from unmixed phase space substructures.

We also tried different models for the vertical velocity dispersion as a function of height. Another functional that gave rise to consistent results (in terms of agreement between different spatial regions and between data sample) was a quartic polynomial. These fits to the velocity data were somewhat worse in terms of $\chi^2$-value and by-eye inspection, but we saw very similar results for the inferred gravitational potential, leading to the same general conclusions.

The uncertainties associated with the velocity distribution are dominant in terms of the inferred gravitational potential. These uncertainties are propagated to the global fits of the exponential function and the simple arm models (see Sect.~\ref{sec:results}).

\section{Results}\label{sec:results}

Our dynamical mass measurement is likely the most robust for the potential difference of roughly $\Phi(400~\pc)-\Phi(0~\pc)$. Between these heights, the stellar number density ratio is roughly 10--15~\% for the three brighter data samples, and 30~\% for the dimmest data sample. Furthermore, the spiral arms are a thin disk phenomenon; staying closer to the Galactic plane makes the measurement more local and will better resolve the spiral arms, as the gravitational potential at greater heights is influenced by a larger area of the Galactic disk. 
This potential difference is a good proxy for and closely proportional to the thin disk surface density, which is discussed in Sect.~\ref{sec:jeans}.

\begin{figure*}
    \centering
    \includegraphics[width=0.85\textwidth]{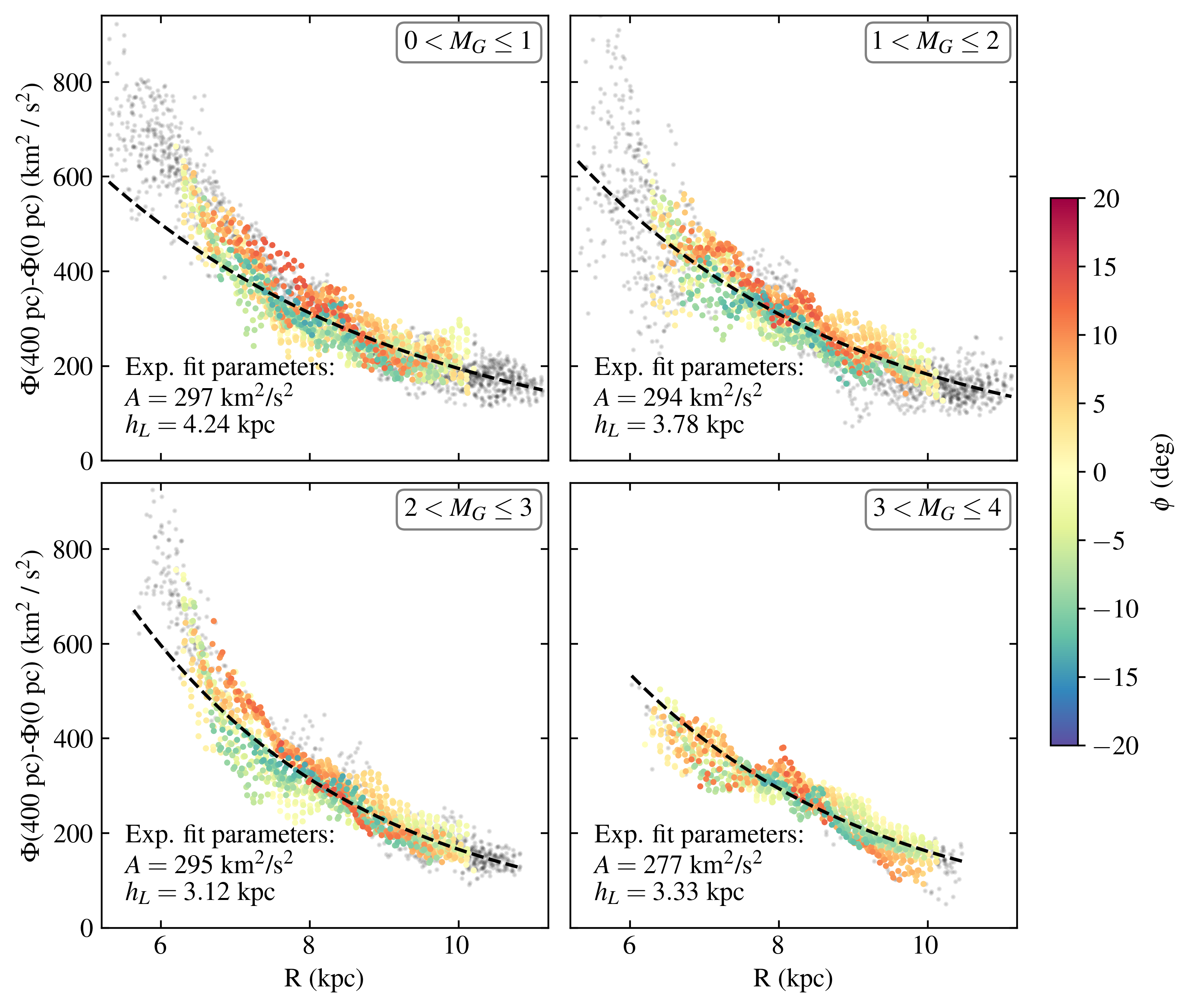}
    \caption{Gravitational potential, $\Phi(400~\pc)-\Phi(0~\pc)$, as a function of Galactocentric radius, for the four stellar samples (as labelled in the panels' top right corner). The dashed line is the exponential function of Eq.~\eqref{eq:exp_func}, fitted to data points with $\sqrt{X^2+Y^2}<2~\kpc$, which excludes distant and dubious spatial regions. The scatter points correspond to area cells in the $(X,Y)$-plane. Area cells that are included in the exponential function fit are coloured according to their Galactocentric disk plane angle ($\phi$); excluded data cells have light grey and smaller scatter points. The colour bar and axis ranges are shared between all panels.}
    \label{fig:all4_scatter}
\end{figure*}

Figure~\ref{fig:all4_scatter} shows our inferred values of $\Phi(400~\pc)-\Phi(0~\pc)$ as a function of Galactocentric radius, for all four data samples. The markers each correspond to an area cell, where the colour denotes their Galactocentric angle $\phi$.
Coherent structure visible in the marker colours (denoting Galactocentric angle $\phi$) reveal non-axisymmetric structure in the inferred surface density results; for example, at the Solar radius ($R=8.2~\kpc$) the results for positive $\phi$ are over-dense compared to negative $\phi$, in all four data samples. We also show an exponential function fitted separately to each data sample, which takes the form
\begin{equation}\label{eq:exp_func}
    f(R) = A \times \exp \left( \frac{R-R_\odot}{h_L} \right),
\end{equation}
where $A$ is the Solar radius value, and $h_L$ is the disk scale length parameter. The data samples that are more distant ($\sqrt{X^2+Y^2}>2~\kpc$) are likely less reliable and therefore masked in this fit. The fitted exponential function parameter values are listed in the respective panels of Fig.~\ref{fig:all4_scatter} and they agree well between the four data samples. We show only the best fit value; the statistical error is small and systematic uncertainties dominate.

\begin{figure*}
    \centering
    \includegraphics[width=0.85\textwidth]{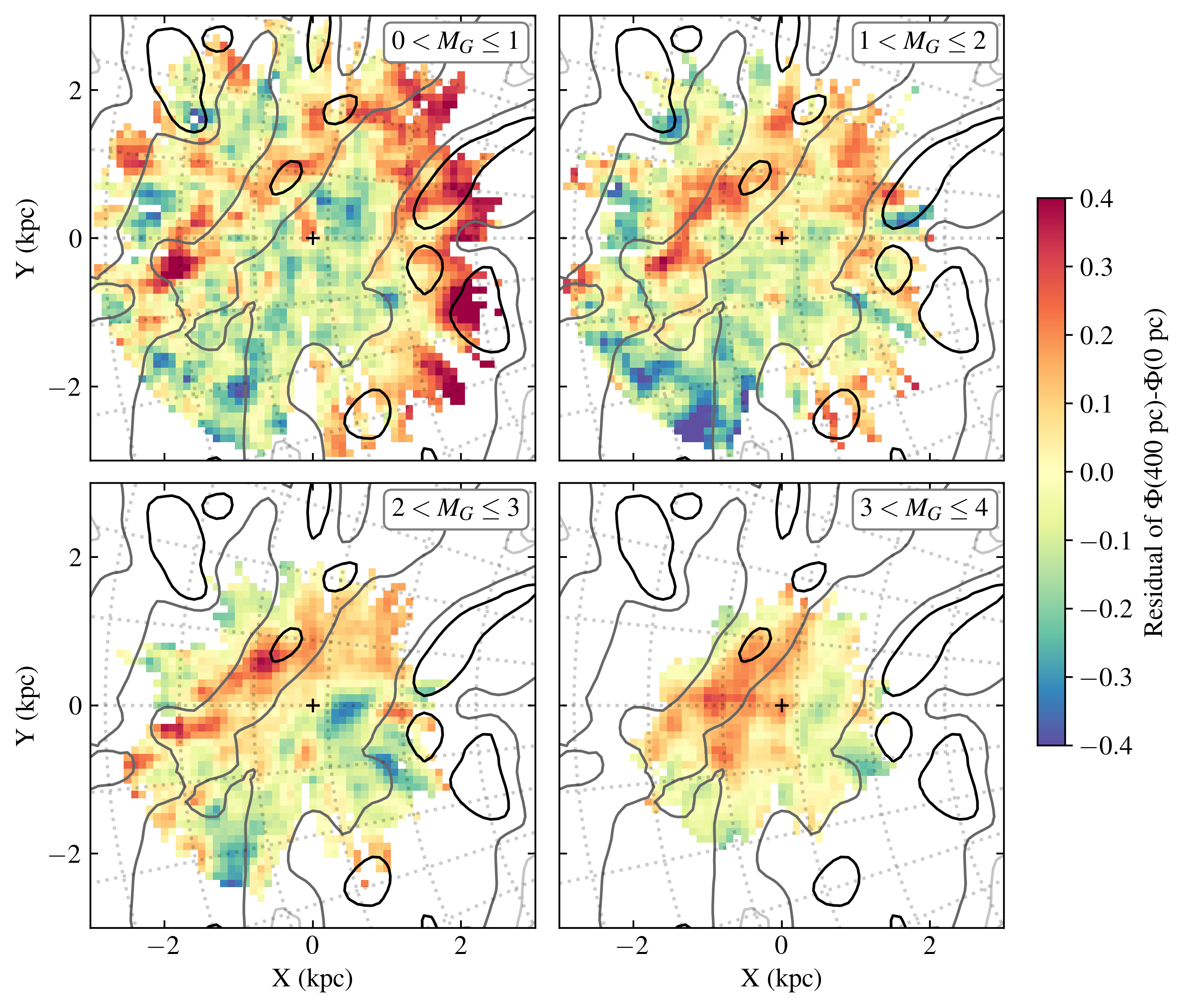}
    \caption{Residual of $\Phi(400~\pc)-\Phi(0~\pc)$ with respect to a fitted exponential, projected on the $(X,Y)$-plane, where each panel corresponds to one of our four data samples. In each panel, dotted grey lines show Galactocentric iso-radial and iso-azimuthal lines, and the black plus sign in the center marks the Solar position. The grey contour lines correspond to the spiral arms as mapped by \cite{Poggio2021}, as seen in Fig.~\ref{fig:poggio}.}
    \label{fig:all4_disk_residuals}
\end{figure*}

\begin{figure}
    \centering
    \includegraphics[width=1.\columnwidth]{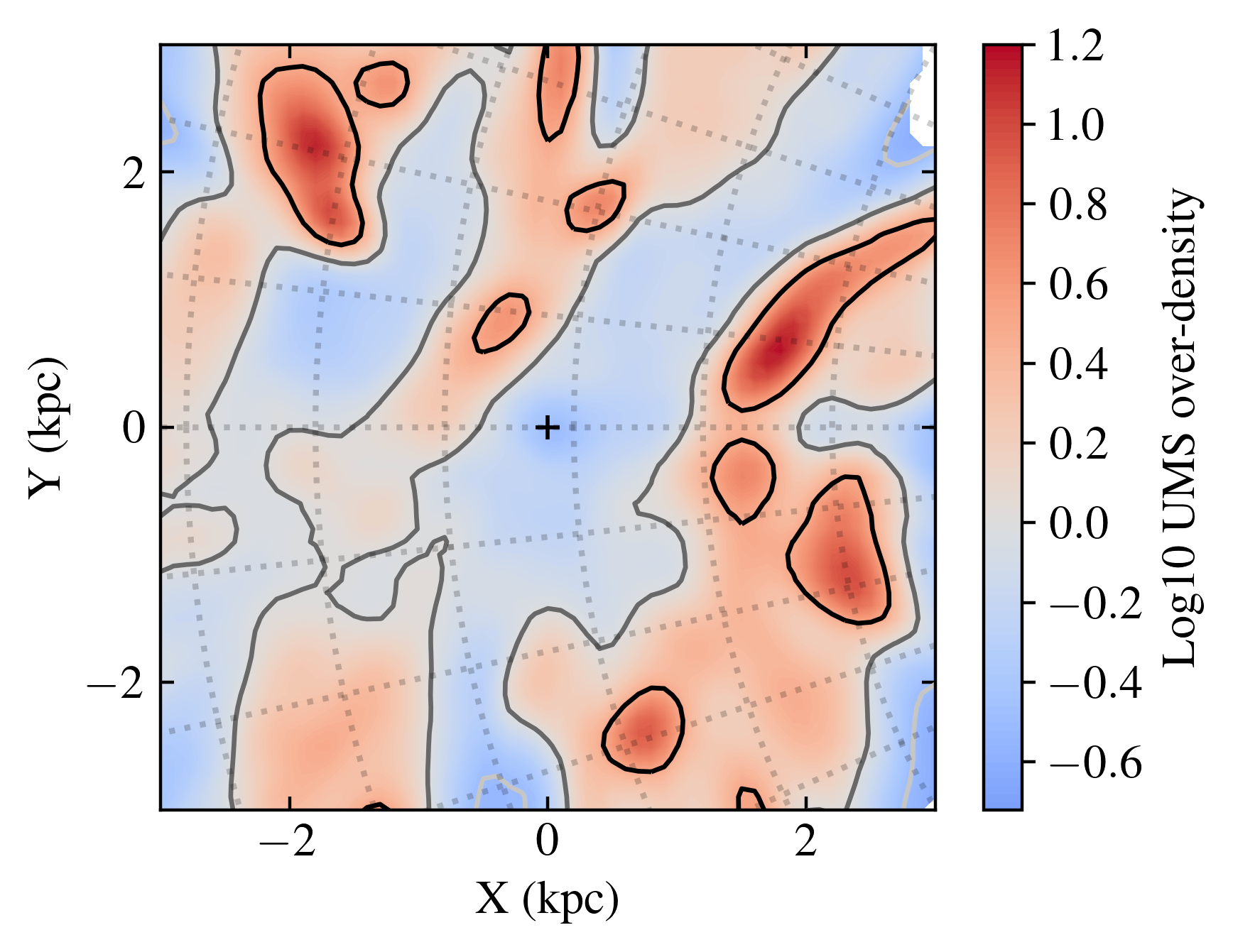}
    \caption{Over-density of upper main sequence stars in the $(X,Y)$-plane, as a tracer of spiral arms. Results are taken directly from \citet{Poggio2021}. The grey-scale contour lines are the same as in Fig.~\ref{fig:all4_disk_residuals}, corresponding to values $(-0.5,\, 0,\, 0.5)$.}
    \label{fig:poggio}
\end{figure}

The residual of the inferred $\Phi(400~\pc)-\Phi(0~\pc)$ as compared to the fitted exponential function is shown in Fig.~\ref{fig:all4_disk_residuals}, for all four data samples. The residual of an area cell is defined according to
\begin{equation}
    \text{residual} = \frac{\text{measurement}-f(R)}{f(R)},
\end{equation}
where $f(R)$ is the fitted exponential function. The residuals of the four data samples show similar results. Most notably, they have the same over-dense structure, stretching roughly from $(X,Y) = (0, 1.5)~\kpc$ to $(X,Y) = (-1.5, -0.5)~\kpc$. This feature agrees well with the Local Arm revealed in the map of young stars by \cite{Poggio2021}, shown in detail in Fig.~\ref{fig:poggio} and as grey contour lines in Fig.~\ref{fig:all4_disk_residuals}.

In order to make quantitative statements about the over-density that we identify with the Local Arm, we fit a simple analytic model to the residual seen in Fig.~\ref{fig:all4_disk_residuals}. We model the ridge of the over-density as a 1d curve in the Galactic disk plane, as a logarithmic spiral of the form
\begin{equation}\label{eq:tilde_R}
    \tilde{R}(\phi) = \tilde{R}_0 \exp \left[ -\phi \tan(\tilde{\alpha}) \right],
\end{equation}
where $\tilde{R}_0$ is the Galactocentric radius where the arm crosses the $X$-axis (i.e., $\phi=0$), and $\tilde{\alpha}$ is the pitch angle. The relative density in this model is given by the disk plane projected distance to the ridge, according to
\begin{equation}\label{eq:tilde_over-dens}
    \text{residual}(R,\phi) = 
    \tilde{a} + \tilde{b} \exp \left\{ -\frac{[R-\tilde{R}(\phi)]^2 \cos^2(\tilde{\alpha})}{2 \tilde{L}^2} \right\},
\end{equation}
where $\tilde{a}$ and $\tilde{b}$ are the residual base-line and spiral arm amplitudes, and $\tilde{L}$ is the spiral arm width.

\begin{figure*}
    \centering
    \includegraphics[width=0.85\textwidth]{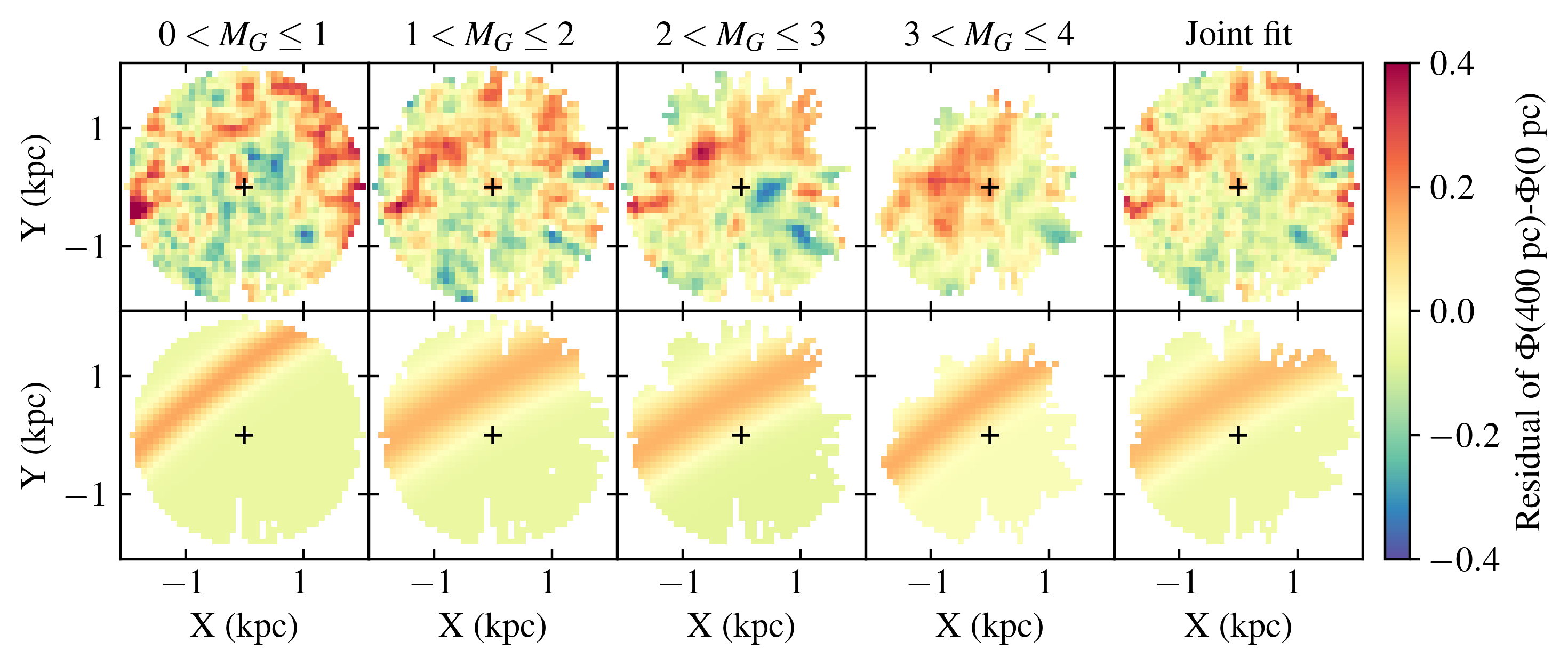}
    \caption{Simple spiral arm model, fitted to the residual of $\Phi(400~\pc)-\Phi(0~\pc)$, for non-masked area cells within a distance of 2~kpc.}
    \label{fig:fitted_arms}
\end{figure*}

The residuals have some strong outliers; for this reason, we use the 1-norm in our fit (minimising $|\text{data}-\text{model}|$, instead of $|\text{data}-\text{model}|^2$), which also gives somewhat more consistent results between data samples. Additionally, we perform a joint fit of all four data samples, which is probably more robust and less prone to over-fitting. The fitted models are shown in Fig.~\ref{fig:fitted_arms}, and the parameter values are listed in Table~\ref{tab:arm_params}.
The relative amplitudes of the arm over-density fall in the range of 18--24~\%, while the joint fit gives 20~\%. The fitted pitch angle is roughly 50$^\circ$--60$^\circ$, which is large compared to typical values in the literature \citep{Levine2006,Vallee2005,Vallee2017}. However, our value is likely only applicable to the very local area of study (e.g., similar to the high pitch angle segment of the Sagittarius Arm analysed by \citealt{Kuhn2021}). The arm width of the joint fit is 0.4~kpc.

\begin{table*}
\caption{Fitted parameters of the simple spiral arm over-density model, as described in Eqs.~\eqref{eq:tilde_R} and \eqref{eq:tilde_over-dens}, for each of the four data samples. The final row is a combination of the $\tilde{a}$ and $\tilde{b}$ parameters, and corresponds to the peak relative over-density as compared to the disk areas far from the spiral arm.}
\label{tab:arm_params}
\centering
\begin{tabular}{{c | c c c c c c}} 
& $\tilde{\alpha}$ (deg) & $\tilde{R}_0$ (kpc) & $\tilde{L}$ (kpc) & $\tilde{a}$ & $\tilde{b}$ & $\tilde{b}/(1+\tilde{a})$ \\
\hline
$0 < M_G \leq 1$ & 48.2 & 9.77 & 0.26 & --0.057 & 0.224 & 0.24 \\
$1 < M_G \leq 2$ & 57.0 & 9.81 & 0.44 & --0.059 & 0.206 & 0.22 \\
$2 < M_G \leq 3$ & 59.1 & 9.52 & 0.43 & --0.075 & 0.224 & 0.24 \\
$3 < M_G \leq 4$ & 52.5 & 9.24 & 0.30 & --0.017 & 0.173 & 0.18 \\
Joint fit & 57.0 & 9.58 & 0.38 & --0.053 & 0.190 & 0.20 \\
\end{tabular}
\end{table*}

\section{Discussion}\label{sec:discussion}

Using vertical Jeans analysis and state-of-the-art data processing, we have produced 2d maps of the thin disk surface density within a distance of 3~kpc of the Sun. Overall, the results of our four data samples agree well. We see the same general over-dense feature in all data samples, and they also agree well with spiral arm tracers based on stellar age and metallicity.

We fitted a simple axisymmetric model to our results, as defined in Eq.~\eqref{eq:exp_func}. The inferred values for normalisation constant (parameter $A$) agree well with previous results for the immediate Solar neighbourhood \citep{Schutz2018,Widmark-sub-regions-DR2,Widmark-spiral-II}. The inferred thin disk scale lengths (parameter $h_L$) fall in range 3.3--4.2~kpc. This quantity varies in previous studies: a review from 2016 by \cite{BlandHawthorn2016} summarizes 15 articles to give $2.6\pm 0.5~\kpc$; later analyses include $2.2\pm 0.1~\kpc$ \citep{Widmark-spiral-III}, $2.4 \pm 0.1~\kpc$ \citep{Wang2022}, roughly 3.9~kpc \citep{Robin2022}, and $2.17^{+0.18}_{-0.08}~\kpc$ \citep{Ibata2023}. An explanation for these discrepancies could be variations in the studied tracer samples and spatial volumes; this could give rise to different results if axisymmetry is broken or if the disk scale length is not constant with respect to $R$. In our own results, the masked data cells at low Galactocentric radii ($R\lesssim 7~\kpc$) seem to follow a steeper profile, likely better described by a broken exponential function. However, we refrain from making a more definite or quantitative statement about this, since our results are quite discrepant beyond 2~kpc and potentially affected by systematic biases, especially towards the Galactic center. The data sample with $1 < M_G \leq 2$ seems particularly discrepant at low $R$; this data sample has the lowest number count, which likely makes it more prone to systematic errors (how such errors operate is very complicated considering the underlying data processing and how it is affected by low statistics, e.g., the spectro-photometric distances affected by dust extinction, which is especially severe at low $R$).

We have detected the Local Arm in our thin disk surface density measurements, as revealed by the residual maps in Fig.~\ref{fig:all4_disk_residuals}, and we can compare and distinguish between other recent probes of the Local Arm morphology. The Local Arm model by \cite{Reid2019}, based on masers as a tracer, is somewhat discrepant with recent observations of young star tracers (see figure 5 in \citealt{Poggio2021} for a direct comparison with \citealt{GaiaDrimmel2023,Gaia2023RecioBlanco}). In the former, the Local Arm barely extends into negative $X$; instead, the outer Perseus Arm has a low pitch angle and curves through $(X,Y) \simeq (-2,-2)~\kpc$. In the latter, the Local Arm has a higher pitch angle, at least in the local area at positive $Y$, and also extends into negative $X$ and $Y$, passing through $(X,Y) \simeq (-2,-2)~\kpc$. Our own results favor the latter Local Arm morphology with a higher pitch angle. However, we do not see evidence for a Local Arm over-density that continues all the way to $(X,Y) \simeq (-2,-2)~\kpc$, as traced by \cite{Poggio2021}, although we repeat that our results could be less accurate in spatial volumes beyond roughly 2~kpc.

There are some discrepancies in the inferred surface density of the four data samples, best seen in Fig.~\ref{fig:all4_disk_residuals}, in addition to the general arm-like structure. This noise is larger than can be accounted for by statistical uncertainties, which is indicative of some uncontrolled systematic errors. For example, there are small outlier ares, for example close to $(X,Y)\simeq (-2,-0.4)~\kpc$ for the brightest data sample, and the arm-like over-density is closer to the solar position for the dimmest data sample. Systematic errors could arise in the data or in the data processing itself; for example, there is a very complicated interplay with the dust extinction field and the spectro-astrometric distance estimates from \texttt{StarHorse}.

There could also be more physical reasons that could bias our results, having to do with the dynamical properties of the respective data samples. For example, the dimmest data sample has a greater spatial scale height and is kinematically hotter; around the solar position, its $u$ and $v$ velocity dispersions are 10--25~\% higher, and its vertical velocity dispersion is 20--50~\% higher, as compared to the other three data samples. These properties make the data sample sensitive to a larger area of the disk plane's mass density distribution, and also affects its sensitivity to biases that can arise from a broken equilibrium assumption; for example, the response of a stellar population with respect to a disk vibrational mode can depend on the scale height and kinematic temperature of that stellar population. In fact, \citet{WidmarkGP} found evidence of a breathing mode associated to the local spiral arm (see their figures 9 and 10, for stars with $|z|<300~\pc$), seen in both the stellar number density and vertical velocity distributions; they found a contracting breathing mode (net motion towards the mid-plane) in the inner part of the arm, offset by roughly a quarter wavelength with respect to the arm's stellar over-density, which is indicative a breathing mode that is travelling against the direction of Galactic rotation. \update{In Appendix~\ref{app:biases}, we quantify the possible bias that can arise from such time-varying phase-space structures. In particular, we investigate vertical breathing modes and the phase-space spiral. We find that the former is most significant, but that its relative bias to the inferred value of $\Phi(400~\pc)-\Phi(0~\pc)$ is limited to a few percent. For the latter, its effect is smaller than one percent. In summary, the biases related to such time-varying structures could be sizeable, but far from large enough to explain the arm-like over-density that we observe in this work.}

Despite the systematic errors discussed above, our general results are robust. A scenario where the detected Local Arm over-density is caused by data systematics seems extremely contrived, since it is observed over a range of viewing angles, distances, and across four independent data samples. 
Furthermore, the data samples are quite different in age and small-scale structure; especially the stars of the dimmest data sample are older, kinematically hotter and more smoothly distributed, and do not trace the spiral arms (at least not in a significant manner). The detection of the spiral arm imprint in these data samples is thus not because of the stellar number counts themselves, but because of the gravitational influence that the spiral arms have on the stellar motions.

\section{Conclusion}\label{sec:conclusion}

We have applied the vertical Jeans equation to the Milky Way disk, dividing the disk plane into a square grid with a grid spacing of 100 pc, in order to study the thin disk surface density and its spatial dependence along the Galactic plane. We reach distances of a few kilo-parsecs, enabled by state-of-the-art data processing, utilising spectro-photometric distance measurements, GP regression on the stellar number density field, and radial velocity predictions from Bayesian Neural Networks. Specifically, we measure the gravitational potential difference between the disk mid-plane and a height of 400~pc, which serves as a close proxy for the thin disk surface density.

For the first time, we see dynamically measured evidence for an over-density associated with a nearby spiral arm. This over-density is detected in all of our four data samples (defined by cuts in absolute magnitude), with a relative amplitude of roughly 20~\%. Our results agree well with the Local Arm as traced by stellar metallicity and age, especially those of \cite{Poggio2021}. This constrains models of spiral arms and paves the way for future dynamical mass measurements of the Milky Way spiral structure.

\begin{acknowledgements}
We thank Yassin-Rany Khalil and Giacomo Monari for useful discussions and input. APN is supported by an Early Career Fellowship from the Leverhulme Trust. AW is sponsored by the Swedish Research Council under contract 2022-04283.

This work has made use of data from the European Space Agency (ESA) mission \textit{Gaia} (\url{https://www.cosmos.esa.int/gaia}), processed by the \textit{Gaia} Data Processing and Analysis Consortium (DPAC, \url{https://www.cosmos.esa.int/web/gaia/dpac/consortium}). Funding for the DPAC has been provided by national institutions, in particular the institutions participating in the \textit{Gaia} Multilateral Agreement.

This research utilised the following open-source Python packages: \textsc{Matplotlib} \citep{matplotlib}, \textsc{NumPy} \citep{numpy}.

For the purpose of open access, the author has applied a Creative Commons Attribution (CC BY) licence to any Author Accepted Manuscript version arising from this submission.
\end{acknowledgements}

\bibliographystyle{aa} 
\bibliography{lib.bib} 

\begin{appendix} 


\section{Biases from time-varying structures}\label{app:biases}

\update{
It is well established that there are time-varying structures in the disk, such as the recently discovered phase-space spiral \citep{antoja2018}. When applying the vertical Jeans equation under the assumption of equilibrium, as we do in this work, such time-varying structures can bias the inferred gravitational potential. In this section we discuss and place upper limits to the relative amplitude of such systematic biases.

In order to investigate the possible bias that arises in the presence of a breathing mode or phase-space spiral, we first assume an underlying model for the gravitational potential and stellar tracer population, which we can then perturb. For the gravitational potential, we use the model of baryonic densities from \cite{Schutz2018}, building on the model from \cite{McKee2015}, which also agrees well with the results of this work and other recent dynamical mass measurements of the immediate solar neighbourhood \citep{Buch2019,Salas2021,Widmark-spiral-II}. We use a stellar tracer population which is built from a superposition of three iso-thermal components, according to
\begin{equation}
    f(z, w) =
    a_i \times \frac{\exp\left( - \dfrac{\Phi(z)+w^2/2}{\sigma_{w,i}^2}\right)}{\sqrt{2 \pi \sigma_{w,i}^2}},
\end{equation}
where $a_i = \{0.4, 0.4, 0.2\}$ are the mid-plane amplitudes of the three components, and $\sigma_{w,i} = \{6, 14, 20\}~\kmsec$ are their respective velocity dispersions. The phase-space distribution at all heights is actually fully described by a function that only depends on vertical energy: $f(E_z)$, where $E_z=\Phi(z)+w^2/2$, the nominator in the exponential above (see \citealt{Widmark2019} for more details on this analytic description).

The phase-space distribution is shown in Fig.~\ref{fig:bias_model}. It is consistent with the general shape of the stellar tracer populations we use in this work, both in terms of the vertical velocity variance and stellar number density profile. The precise shape of phase-space distribution of this underlying model is not very important in terms of the arising bias. We tried a range of different tracer populations, with varying scale heights and vertical velocity variance profiles, but saw similar relative biases in the inferred value of $\Phi(400~\pc)-\Phi(0~\pc)$.

\begin{figure*}
\centering\includegraphics[width=0.85\textwidth]{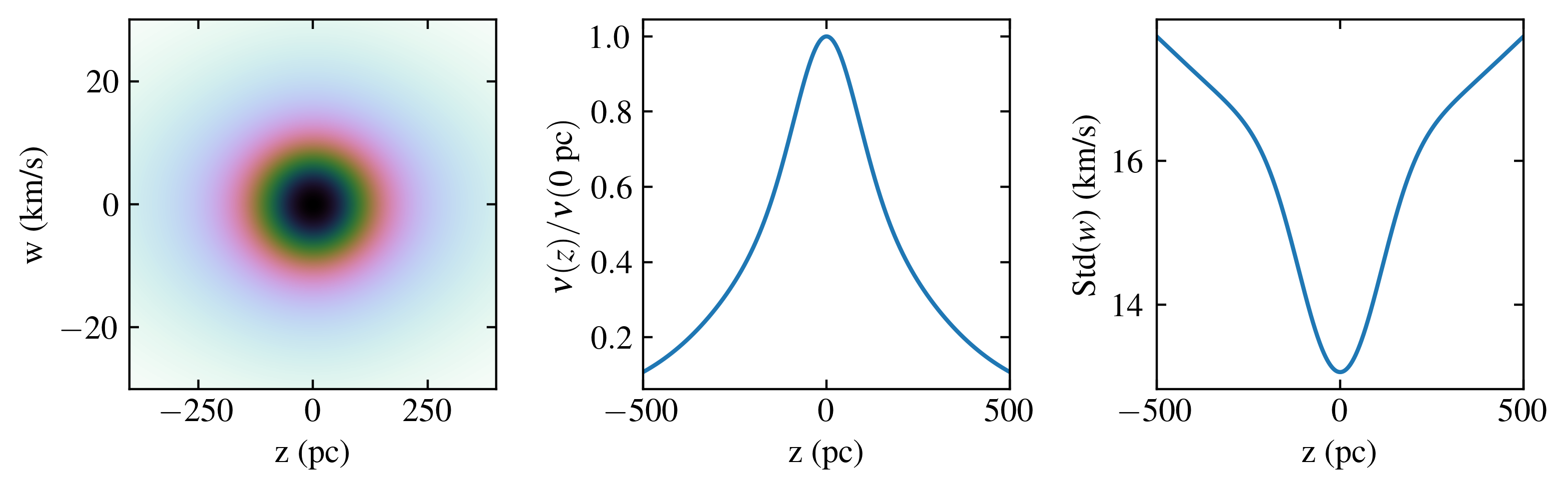}
    \caption{Phase-space distribution of our non-perturbed, steady state model. The left panel shows a 2d histogram in the $(z,w)$ phase-space plane. The middle panel shows the stellar number density profile, normalised to unity in the mid-plane. The right panel shows the standard deviation of vertical velocity, as a function of height.}
    \label{fig:bias_model}
\end{figure*}

\subsection{Breathing mode}

First we consider a breathing mode. We apply a perturbation to the phase-space distribution as a function of vertical energy ($E_z$) and vertical action angle ($\theta$), according to \citep{mathur1990,weinberg1991,widrow2015}
\begin{equation}
    f(E_z, \theta) = f_0(E_z) \{1 + \epsilon E_z\cos [2(\theta - \theta_0] \},
\end{equation}
where $f_0(E_z)$ is the unperturbed distribution, and $\theta_0$ and $\epsilon$ are constants.

We set $\epsilon$ to a value such that $\Delta w_{400~\pc}$ (the difference in mean velocity 400~pc above and below the mid-plane) can be at most $1~\kmsec$, which happens for $\theta_0=\pi/4$. In \cite{GaiaDrimmel2023}, they compare the mean vertical velocity above and below the mid-plane, stating that $|\Delta w|$ reaches values of 2--4~$\kmsec$. However, such large values are seen only in more distant regions of the disk (beyond 3~kpc, see their figure 23). For the more nearby solar neighourhood, similar to the volume we are probing, $|\Delta w|$ is limited to less than $1~\kmsec$. \cite{WidmarkGP} did a similar analysis of the four specific data samples studied in this work, and with a more refined binning in height (see their figures 8, 9 and 10, and appendix). They evaluated $\Delta w$ specifically for the range of $|z| \in [300,500]~\pc$, and saw $|\Delta w|$ typically below $1~\kmsec$, although there are some smaller regions with absolute values of roughly 2~$\kmsec$, especially around $(X,Y)=(0, 1.5)~\kpc$.

The relative bias that arises from the breathing mode with respect to the non-perturbed phase-space distribution, as a function of the angle $\theta_0$, is shown in Fig.~\ref{fig:bias_breathing}. As illustrated in the top panels, the phase-space distribution is the most spatially compressed at $\theta_0 = 0$, which is where the bias is positive (i.e. the inferred potential is over-estimated). Conversely, it is the most spatially stretched out for $\theta_0 = \pi / 2$, where the bias is negative. This breathing mode is fully uniform, in the sense that particles of the same action angle are affected in the same manner. In the real Galactic disk, such an idealised mode would phase-mix. This would likely lower the dynamical mass measurement bias over time, even if the amplitude of $\Delta w_{400~\pc}$ is sustained.

In summary, breathing modes in the stellar disk can likely bias the inferred value of $\Phi(400~\pc)-\Phi(0~\pc)$ by a few percent. The arm-like structure we observe in this work has a relative over-density of roughly 20~\%. To fully explain this as a bias arising due to a breathing mode would require amplitudes corresponding to $\Delta w_{400~\pc} \gtrsim 6~\kmsec$, which we can confidently rule out.

\begin{figure*}
    \centering
    \includegraphics[width=0.85\textwidth]{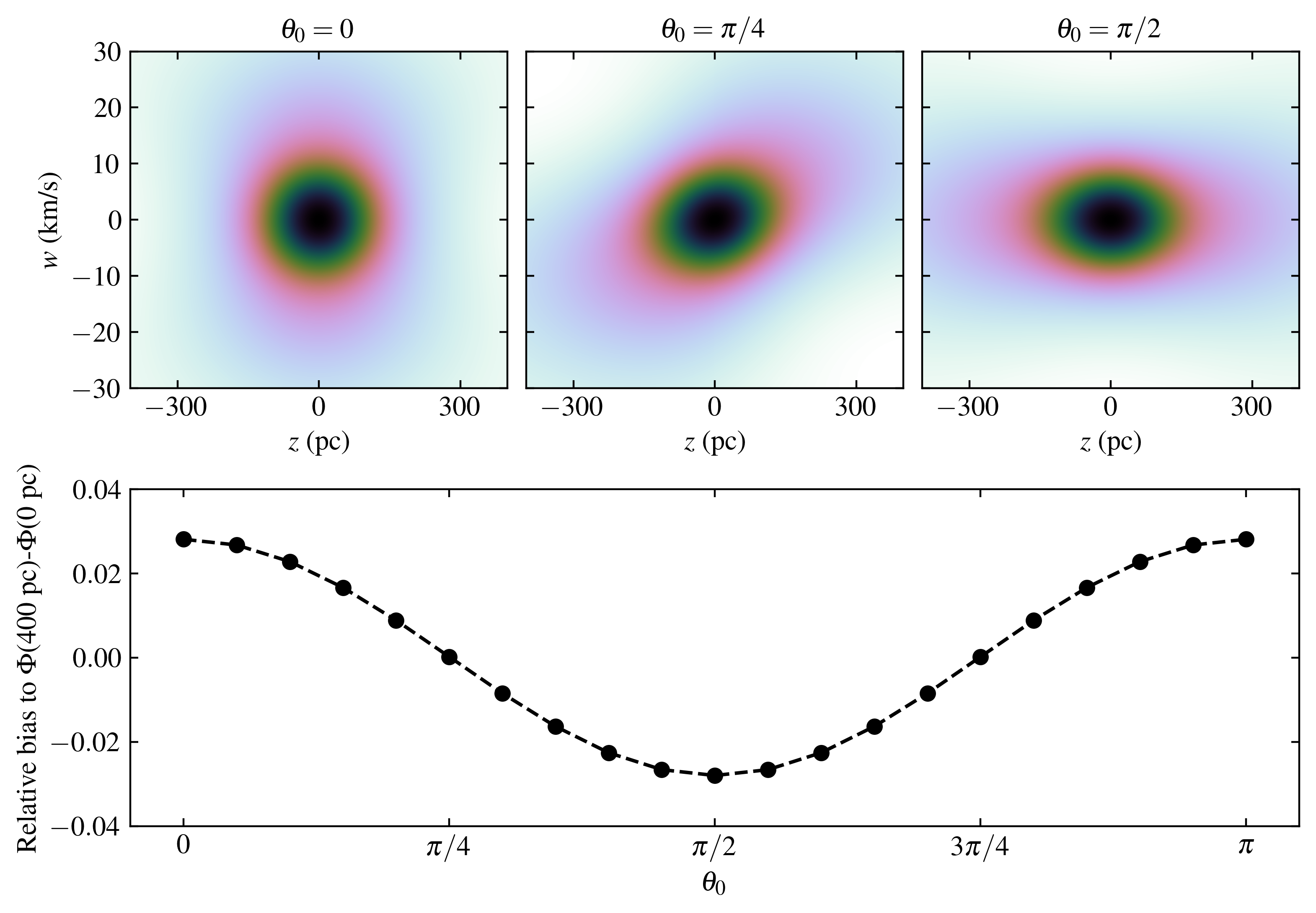}
    \caption{Relative bias to the inferred value of $\Phi(400~\pc)-\Phi(0~\pc)$ using Jeans analysis under the assumption of equilibrium, in the presence of a breathing mode as a function of $\theta_0$. The bias is shown in the bottom panel, where the breathing mode is normalised such that the difference in mean value of $w$ between 400 pc above and below the midplane is $\Delta w_{400~\pc}=1~\kmsec$ when $\theta_0 = \pi/4$. The top panels show how the breathing mode phase, $\theta_0$, affects the phase-space distribution. The perturbation is strongly enhanced in the top panels, for better visibility.}
    \label{fig:bias_breathing}
\end{figure*}

\subsection{Phase-space spiral}

We do a similar test for a phase-space spiral. We use the model from \cite{Widmark-spiral-I}, with a relative over-density amplitude of 20~\% and a time of 500~Myr since the perturbation was produced. These parameter values give rise to a phase-space spiral which is qualitatively similar to the one observed in the actual solar neighbourhood. We tried different parameter choices, giving rise to, for example, different degrees of spiral winding, and saw similar results in terms of the strength of the possible bias.

The relative bias with respect to the non-perturbed phase-space distribution is shown as a function of $\theta_0$ in Fig.~\ref{fig:bias_spiral}. Here, $\theta_0$ is in range $[0,2\pi]$, since it is an asymmetric structure (as opposed to the symmetric breathing mode of Fig.~\ref{fig:bias_breathing}). Here, the bias to $\Phi(400~\pc)-\Phi(0~\pc)$ is limited to less than a percent.}

\begin{figure*}
    \centering
    \includegraphics[width=0.85\textwidth]{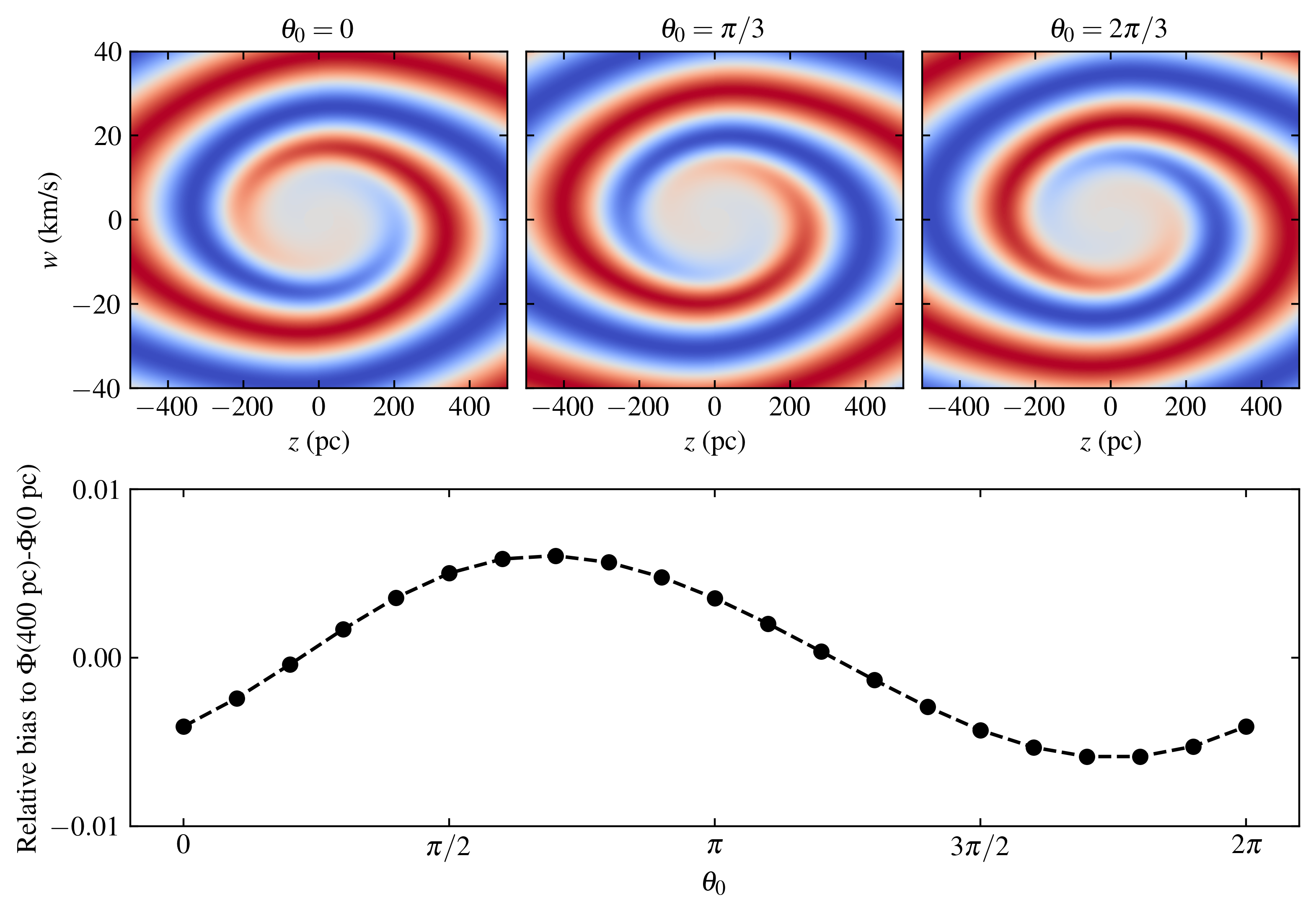}
    \caption{Relative bias to the inferred value of $\Phi(400~\pc)-\Phi(0~\pc)$ using Jeans analysis under the assumption of equilibrium, in the presence of a phase-space spiral as a function of $\theta_0$. The top panels show the relative perturbation of the spiral (unlike Fig.~\ref{fig:bias_breathing}, which showed the total number densities), where red (blue) correspond to a over-density (under-density), whose relative amplitude is in range $[-20,20]~\%$.}
    \label{fig:bias_spiral}
\end{figure*}

\end{appendix}

\end{document}